\documentclass[a4paper,11pt]{article}
\pdfoutput=1 

\usepackage{jheppub} 

\usepackage[T1]{fontenc} 

\usepackage{braket} 

\usepackage{dsfont} 

\newcommand{\tr}{\ensuremath{\mathrm{tr}}}
\newcommand{\Res}{\mathrm{Res}}
\renewcommand{\vec}{\mathbf}

\title{\boldmath  On equivalent methods for functional determinants}

 \author[a,b]{Matthias Carosi}
 \affiliation[a]{Physik Department T70, Technische Universit\"at M\"unchen,\\James-Franck-Str., D-85748 Garching, Germany}
 \affiliation[b]{International Centre for Theoretical Physics Asia-Pacific (ICTP-AP), \\University of Chinese Academy of Sciences, 100190 Beijing, China}

\emailAdd{matthias.carosi@tum.de}

\arxivnumber{2601.08686}

\abstract{
Computing functional determinants of differential operators is central to any field-theoretical calculation relying on a saddle-point expansion.
A variety of approaches is available for the computation that avoid having to know the eigenspectrum of the operator, and in particular the Gel'fand-Yaglom theorem and the Green's function method.
In this note, we show how both approaches can be constructed using a contour integral argument and conclude that these are completely equivalent for computing ratios of determinants of one-dimensional operators.
Furthermore, we comment on the presence of vanishing as well as negative eigenvalues and show how the Green's function method provides a natural prescription for handling them.
}

\begin{document} 
\begin{flushright}
	TUM-1590/26\\ 
    Published in \href{https://doi.org/10.1007/JHEP06(2026)073}{JHEP 06 (2026) 073}
\end{flushright}
\maketitle
\flushbottom

\section{Introduction}
\label{sec:intro}
Functional determinants arise naturally in many computations in field theory.
Specifically, they appear as the next-to-leading-order (NLO) term in the saddle-point expansion of a Euclidean path integral\,\cite{Ramond:1981pw,Coleman:1985rnk,Weinberg:2012pjx}.
As an example, let us consider the theory of a single scalar field~$\phi$ described by the Euclidean action $S_{\mathrm{E}}[\phi]$.
We can compute the Euclidean path integral as a saddle-point expansion around some configuration $\phi_0$, which solves the Euclidean equation of motion $S_{\mathrm{E}}'[\phi_0]=0$. We define $\phi=\phi_0+\sqrt{\hbar}\eta$, where $\hbar$ is a small parameter and $\eta$ is a fluctuation around the background $\phi_0$.
Expanding for small $\hbar$, we obtain
\begin{align}
    \int \left[ \mathcal{D}\phi\right] e^{-\frac{1}{\hbar}S_{\mathrm{E}}[\phi]} = \: & e^{-\frac{1}{\hbar}S_{\mathrm{E}}[\phi_0]} \int \left[\mathcal{D}\eta\right] e^{-\frac{1}{2}\int_{x,y} \eta_x G^{-1}_{\phi_0,xy}\eta_y} \left[ 1 + \mathcal{O}(\hbar)\right] \notag\\
    = \: & e^{-\frac{1}{\hbar}S_{\mathrm{E}}[\phi_0]} \left( \det G^{-1}_{\phi_0,xy} \right)^{-\frac{1}{2}} \left[ 1 + \mathcal{O}(\hbar)\right] \,,
\end{align}
where we have defined the inverse propagator in the background $\phi_0$,
\begin{equation}
    G^{-1}_{\phi_0,xy} = \: \left[\frac{\delta^2S_{\mathrm{E}}[\phi]}{\delta\phi_x\delta\phi_y}\right]_{\phi=\phi_0} \,.
\end{equation}
This brief derivation illustrates how functional determinants enter field theory calculations, but it is merely a sketch, as it overlooks several key aspects, including normalisation, boundary conditions, and zero modes.
Calculations like this appear when computing the contribution of fluctuations around an instanton configuration\,\cite{tHooft:1976snw}, the one-loop corrections to the nucleation rate during a phase transition\,\cite{Coleman:1977py,Callan:1977pt}, or the radiative corrections to the mass of a soliton\,\cite{Dashen:1974ci,Dashen:1974cj}.

While, in principle, one could compute a functional determinant by finding the spectrum of the operator and then evaluating the product of the eigenvalues, this is generally very hard and, therefore, a quicker route is desirable.
Strictly speaking, we usually encounter ratios of functional determinants, which turns out to be a more amenable problem.
To this end, Gel'fand and Yaglom found that the ratio of functional determinants can be recast into an initial-value problem with modified boundary conditions. This result goes under the name of Gel'fand-Yaglom theorem, and it first appeared in Ref.\,\cite{Gelfand:1959nq}.
The theorem has become the most popular approach to computing ratios of functional determinants, and it has been successfully implemented for the study of nucleation at zero\,\cite{Kiselev:1984,Baacke:2003uw,Dunne:2005rt} and finite\,\cite{Baacke:1993ne,Ekstedt:2021kyx,Ekstedt_2023} temperature.

Another, less popular approach to computing the ratio of functional determinants is the Green's function method, also known as the resolvent method.
This approach turns computing the ratio of functional determinants of two operators into finding the respective Green's functions and integrating over their difference. Initially developed for the calculation of the sphaleron rate for baryon relaxation~\cite{Baacke:1993aj,Baacke:1993jr}, the method has been successfully applied to vacuum decay\,\cite{Garbrecht:2015oea} and instanton calculations~\cite{Ai:2024taz}.
A pedagogical presentation of the Green's function method is available in a set of lecture notes by Garbrecht~\cite{Garbrecht_2021}.

In this note, we provide a parallel derivation of the Gel'fand-Yaglom theorem and the Green's function method using contour integral techniques. Our derivation follows closely the techniques used by Kirsten and McKane to prove the Gel'fand-Yaglom theorem~\cite{Kirsten:2003py}, later presented in a set of lecture notes by Dunne~\cite{Dunne_2008}.
The main novelty of this article is that our derivation shows the complete equivalence of the Gel'fand-Yaglom theorem and the Green's function method for computing the ratio of determinants of operators in one dimension.
A constructive argument for this equivalence first appeared in the lecture notes by Garbrecht~\cite{Garbrecht_2021} and is fully complementary to our present derivation.
Furthermore, we clarify some aspects of the Green's function method, specifically in what concerns the treatment of zero and negative eigenvalues.
To close the circle, we include a discussion on the heat kernel approach and show how all three methods are one and the same.

In section~\ref{sec:zeta_regularisation}, we present the conventional definition of the determinant using the $\zeta$-function regularisation. This is the starting point of any manipulation aiming at reducing the computation of the determinant to a simpler problem.
We present the contour integral argument in section~\ref{sec:derivation}, and we show how the Gel'fand-Yaglom theorem and the Green's function method emerge naturally within the same framework.
In section~\ref{sec:green_method}, we present the conventional derivation of the Green's function method based on the spectral decomposition, which is particularly apt for the generalisation to higher dimensions.
The treatment of vanishing eigenvalues is crucial for physics-related applications, and we tackle this in section~\ref{sec:zero-modes} with a focus on the Green's function method.
In section~\ref{sec:concerning_heat_kernel}, we comment on the heat-kernel method and its relation to the Green's function approach.
We present our conclusions in section~\ref{sec:conclusions}.

\section{The $\zeta$-regularisation}
\label{sec:zeta_regularisation}
In this first section, we provide a rigorous definition of functional determinants via the $\zeta$-function regularisation. As we will see, subsequent derivations rely on expressing the $\zeta$-function in some convenient way.

In the following, we will work in only one spatial dimension. In fact, for many physically relevant questions, multi-dimensional problems can be reduced to one-dimensional ones\,\cite{Dunne:2006ct}.
We work with coordinates on the interval
\begin{equation}
	x\in[a,b] \subset \mathbb{R} \,,
\end{equation}
where~$a$ and~$b$ are allowed to be infinite.
We consider a linear second order differential operator~$\widehat{O}$ acting on the Hilbert space~$\mathcal{H}$ endowed with the~$L^2$ scalar product,
\begin{equation}
	\langle f, g \rangle = \: \int_a^b d \mu(x) \, f^*(x) g(x) \,,
\end{equation}
where $\mu(x)$ is a measure.
For example, for radially symmetric problems in $D$ dimensions, $x\in[0,\infty)$ is the radius and $d\mu(x) = x^{D-1}dx$ is the integral measure.
The operator~$\widehat{O}$ is self-adjoint with respect to the scalar product, namely
\begin{equation}
	\langle f, \widehat{O}g\rangle = \: \langle \widehat{O} f , g \rangle \,.
\end{equation}
Here and in the rest of this article, $*$ denotes the complex conjugation.
Furthermore, we assume that the spectrum of~$\widehat{O}$ is fully discrete and bounded from below, namely
\begin{equation}
	\sigma(\widehat{O},\mathcal{H}) = \: \left\{ \lambda_i\right\}_{i\in\mathbb{N}} \subset \: \mathbb{R}\,, \qquad \mathrm{and} \quad \lambda_0 < \lambda_i \quad \forall \quad i>0\,.
\end{equation}
The spectrum $\sigma(\widehat{O},\mathcal{H})$ is the set of eigenvalues of the operator $\widehat{O}$ over the Hilbert space $\mathcal{H}$, namely
\begin{equation}
	\sigma(\widehat{O},\mathcal{H}) = \: \Big\{ \lambda_i \; | \; \widehat{O} \phi_i = \lambda_i \phi_i \quad \mathrm{and} \quad \phi_i \in \mathcal{H} \Big\} \,. 
\end{equation}
Though we use the assumption of a discrete spectrum extensively in the following derivation, the results can be extended to the general case when the spectrum has a continuous part.
The reality of the spectrum is ensured by the operator~$\widehat{O}$ being Hermitian in~$\mathcal{H}$.
The spectrum is found by solving the eigenvalue problem
\begin{equation}
	\label{eqn:eigenvalue_problem}
	\begin{cases}
		(\widehat{O} \phi_i)(x) = \: \lambda_i \phi_i(x) \\
		\phi_i(a) = \: \phi_i(b) = \: 0 
	\end{cases}
	\,,
\end{equation}
and the eigenfunctions form an orthonormal eigenbasis
\begin{equation}
	\langle \phi_i , \phi_j \rangle = \: \delta_{ij}\,.
\end{equation}
Although we have defined the differential problem~\eqref{eqn:eigenvalue_problem} with Dirichlet boundary conditions on the interval, we may be interested in more generic boundary conditions. For example, in the context of vacuum decay, one looks at problems with mixed boundary conditions~$\phi'(a)=\phi(b)=0$.
The generalisation of the present discussion to such boundary conditions is rather straightforward~\cite{Kirsten:2003py,Kirsten:2004qv}.

At this point, we make one more crucial assumption, that all eigenvalues~$\lambda_i$ are non-zero.
In fact, though we are often interested in operators with zero modes, the general strategy to treat them is to work in a Hilbert space where they are not present, as we discuss in section~\ref{sec:zero-modes}.
In practice, this means that the operator~$\widehat{O}$ acting on~$\mathcal{H}$ does not exhibit any eigenmode with a vanishing eigenvalue.

A rigorous definition of the functional determinant of the operator~$\widehat{O}$ can be given in terms of the~$\zeta$-function, defined as
\begin{equation}
	\zeta_{\widehat{O}}(t) = \: \tr \, \widehat{O}^{-t} \,.
\end{equation}
Then, the functional determinant of the operator is defined by the relation
\begin{equation}
	\label{eqn: def det zeta function}
	\det \widehat{O} = \: e^{- \zeta_{\widehat{O}}'(0)} \,,
\end{equation}
where the prime denotes the derivative.
Here and in the following, traces and determinants must always be understood as taken over the space~$\mathcal{H}$.
The~$\zeta$-function associated with the operator~$\widehat{O}$ can be readily written in terms of the spectrum
\begin{equation}
	\zeta_{\widehat{O}} (t) = \: \sum_i \lambda_i^{-t} \,,
\end{equation}
so that its derivative computes to
\begin{equation}
	\zeta_{\widehat{O}} ' (t) = \: - \sum_i \lambda_i^{-t} \log \lambda_i \: \xrightarrow{t\to0} \: - \sum_i \log \lambda_i = \: - \log \prod_i \lambda_i \,,
\end{equation}
and using the definition in Eq.\,\eqref{eqn: def det zeta function}, one finds
\begin{equation}
	\det \widehat{O} = \: \prod_i \lambda_i \,.
\end{equation}
Since there is no concept of product over a continuous set, Eq.\,\eqref{eqn: def det zeta function} serves as the definition for general operators that also have a continuous part of the spectrum.
Thus, we have translated the problem of computing the functional determinant of an operator to obtaining the derivative of the respective~$\zeta$-function at zero.
Next, we show how this can be recast as a differential problem, either through the Gel'fand-Yaglom theorem or the Green's function method.

\section{A contour integral derivation}
\label{sec:derivation}
We start with defining a function~$F_{\widehat{O}}$ analytic on the whole complex plane and such that
\begin{equation}
	F_{\widehat{O}}(\lambda) = \: 0 \,, \qquad \text{if and only if} \qquad \lambda \in \sigma (\widehat{O} ,\mathcal{H} ) \subset \mathbb{R} \,,
\end{equation}
and the zeroes are \emph{simple}.
Then, the~$\zeta$-function associated to the operator~$\widehat{O}$ can be written as a contour integral
\begin{equation}
	\label{eqn:zeta_as_contour_integral}
	\zeta_{\widehat{O}}(t) = \: \frac{1}{2\pi i} \oint_{C_+} d\lambda \, \lambda^{-t}\, \frac{d}{d\lambda} \log F_{\widehat{O}} (\lambda) \,.
\end{equation}
The function~$\lambda^{-t}$ introduces a singularity at the origin, and a branch cut whose location we can choose freely.
As we will argue shortly, the integrand also has simple poles at each zero of~$F_{\widehat{O}}$, namely at each~$\lambda_i\in\sigma(\widehat{O},\mathcal{H})$.
Having chosen~$F_{\widehat{O}}$ to be analytic everywhere, the integral exhibits no other poles or branch cuts.

\begin{figure}
	\centering
	\includegraphics[width=0.6\textwidth]{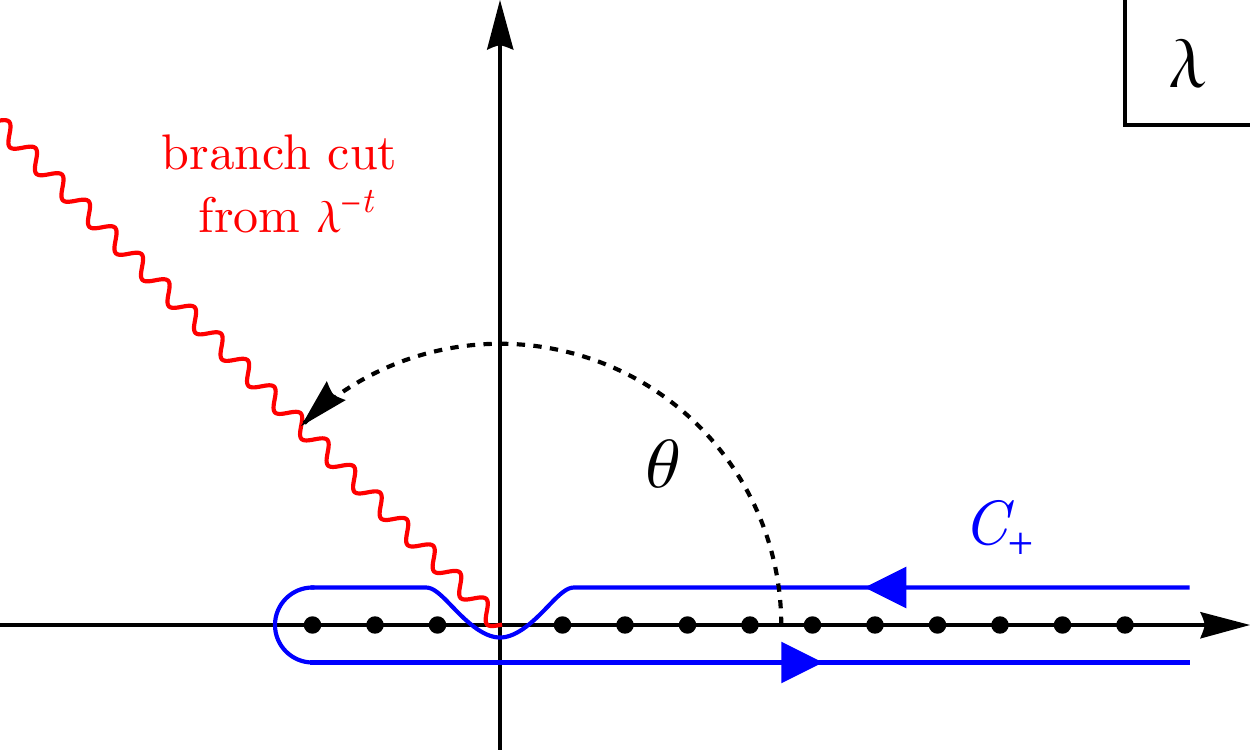}
	\caption[Contour~$C_+$]{The contour~$C_+$ on the complex~$\lambda$-plane. The black dots represent the poles of the integrand, while the red wavy line is the branch cut coming from~$\lambda^{-t}$, located at an angle~$\theta$ with respect to the positive real axis. Adapted from Ref.\,\cite{Kirsten:2004qv}.}
	\label{fig:C+_contour}
\end{figure}

The contour~$C_+$ is then chosen so that it wraps around each of the possibly infinite simple poles and avoids the singularity at zero and the associated branch cut. The latter is chosen to be at an angle~$\theta$ with respect to the positive real axis.
This is represented in Figure~\ref{fig:C+_contour}.

We can evaluate the contour integral in Eq.\,\eqref{eqn:zeta_as_contour_integral} via the residue theorem
\begin{equation}
	\frac{1}{2\pi i} \oint_{C_+} d\lambda \, \lambda^{-t}\, \frac{d}{d\lambda} \log F_{\widehat{O}} (\lambda) = \:
	\sum_{\lambda_i\in\sigma(\widehat{O},\mathcal{H})}
	 \Res_{\lambda =\lambda_i} \, \lambda^{-t}\, \frac{d}{d\lambda} \log F_{\widehat{O}} (\lambda)  \,.
\end{equation}
The residue at each pole is easy to find. First, recall our assumption that~$F_{\widehat{O}}$ has simple zeroes at each~$\lambda_i$ and is analytic everywhere, so that around each zero we can write
\begin{equation}
	F_{\widehat{O}} (\lambda) = c_i (\lambda - \lambda_i) + \mathcal{O}((\lambda-\lambda_i)^{2})\,,
\end{equation}
where $c_i$ is a complex number.
We then compute
\begin{equation}
	\frac{d}{d\lambda} \log F_{\widehat{O}} (\lambda) = \: \frac{F_{\widehat{O}}'(\lambda)}{F_{\widehat{O}}(\lambda)} = \frac{1}{\lambda-\lambda_i} + \mathcal{O}(1) \,.
\end{equation}
We find that the poles at the eigenvalues of~$\widehat{O}$ are all \emph{simple} and with residue one.
Using this, we can then compute the sum over the residues
\begin{equation}
	\sum_{\lambda_i\in\sigma(\widehat{O},\mathcal{H})} \Res_{\lambda =\lambda_i} \, \lambda^{-t}\, \frac{d}{d\lambda} \log F_{\widehat{O}} (\lambda) = \: \sum_{\lambda_i\in\sigma(\widehat{O},\mathcal{H})} \, \lambda_i^{-t} = \: \zeta_{\widehat{O}} (t) \,,
\end{equation}
which proves Eq.\,\eqref{eqn:zeta_as_contour_integral}.

\begin{figure}
	\centering
	\includegraphics[width=0.6\textwidth]{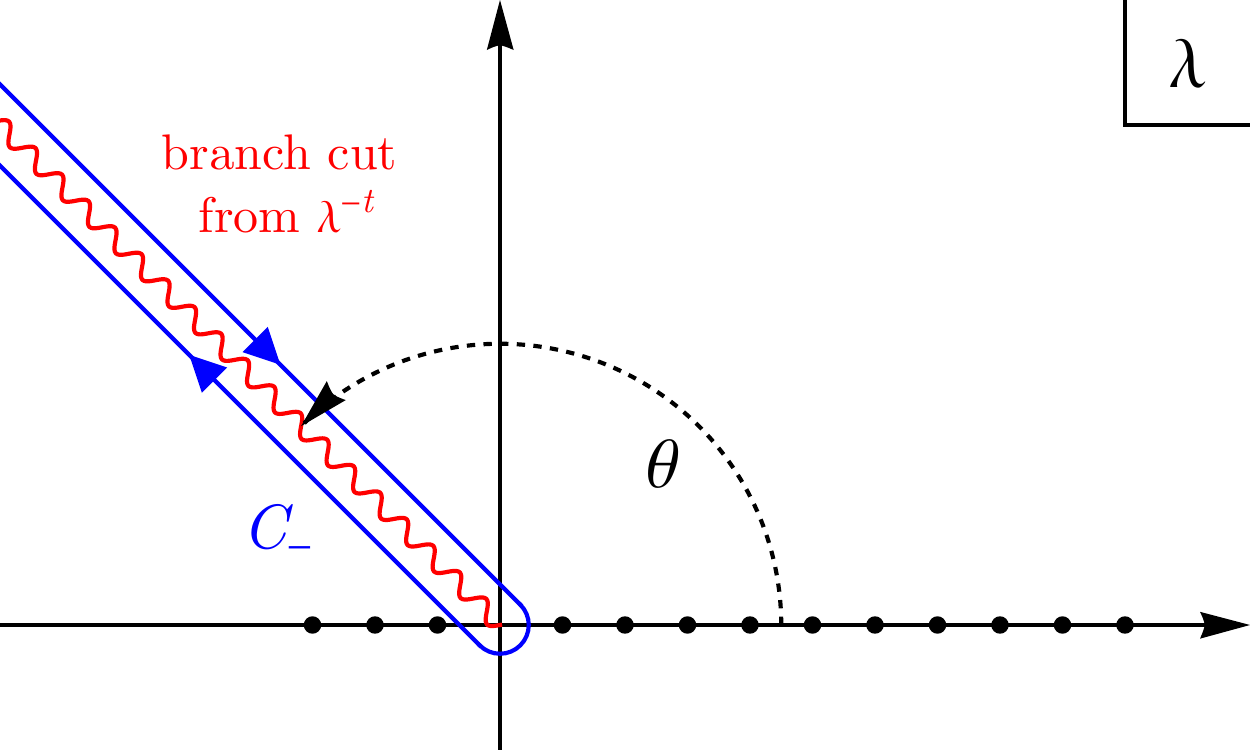}
	\caption[Contour~$C_-$]{The contour~$C_-$ obtained by deforming~$C_+$ as shown in Figure~\ref{fig:C+_contour}. The new contour wraps around the branch cut coming from~$\lambda^{-t}$.
	}
	\label{fig:C-_contour}
\end{figure}

Now that we have an expression for the~$\zeta$-function in terms of a contour integral, we can express the determinant in terms of the function~$F_{\widehat{O}}$.
To do so, we first deform the integration contour from~$C_+$ to~$C_-$, as shown in Figure~\ref{fig:C-_contour}. 
This is allowed because the integrand is analytical everywhere except at the poles and at the branch cut, so that we have
\begin{align}
	\frac{1}{2\pi i} \oint_{C_+} d\lambda \, \lambda^{-t}\, \frac{d}{d\lambda} \log F_{\widehat{O}} (\lambda) = \: & \frac{1}{2\pi i} \oint_{C_-} d\lambda \, \lambda^{-t}\, \frac{d}{d\lambda} \log F_{\widehat{O}} (\lambda) \,.
\end{align}

\noindent
Let us define the integrand
\begin{align}
	h_{\widehat{O}}(t;\lambda) = \: \lambda^{-t} \, \frac{d}{d\lambda} \log F_{\widehat{O}} (\lambda)\,.
\end{align}
When going clockwise around the origin, this function picks up a phase
\begin{align}
	h_{\widehat{O}}(t;e^{- 2\pi i }\lambda) = \: e^{2\pi i t} h_{\widehat{O}}(t;\lambda) \,,
\end{align}
so that the contour integral reduces to 
\begin{align}
	\zeta_{\widehat{O}}(t) = \: & \frac{1}{2\pi i} \oint_{C_-} d\lambda \, h_{\widehat{O}} (t;\lambda) \notag \\
    = \: & \frac{1}{2\pi i} \left\{ \int_{e^{i\theta}\infty}^0 d \lambda \, h_{\widehat{O}}(t;\lambda) + e^{2\pi i t} \int_0^{e^{i\theta}\infty} d\lambda \, h_{\widehat{O}}(t;\lambda)  \right\} \notag \\
	= \: & \frac{e^{2\pi i t} - 1}{2\pi i} \int_0^{e^{i\theta}\infty} d\lambda \, h_{\widehat{O}}(t;\lambda) \,.
\end{align}
The functional determinant is expressed in terms of the derivative of the~$\zeta$-function at zero, which we want to compute next.
Differentiating the integrand with respect to~$t$, we have
\begin{equation}
	\frac{d}{d t} h_{\widehat{O}}(t;\lambda) = \: -(\log\lambda) \, h_{\widehat{O}} (t;\lambda)\,,
\end{equation}
with which we can compute the derivative of the~$\zeta$-function, namely
\begin{align}
	\zeta_{\widehat{O}}'(t) = \: & e^{2\pi i t} \int_0^{e^{i\theta}\infty} d\lambda \, h_{\widehat{O}}(t;\lambda) - \frac{e^{2\pi i t} - 1}{2\pi i} \int_0^{e^{i\theta}\infty} d\lambda \, (\log\lambda) \, h_{\widehat{O}}(t;\lambda) \,.
\end{align}
At this point, we make use of the fact that we are interested in computing a \emph{ratio} of functional determinants. In particular, let us take a second operator~$\widehat{O}_0$ which satisfies all the same assumptions we made for the operator~$\widehat{O}$.
Then, we also have the associated function~$F_{\widehat{O}_0}$, with simple zeroes only at the eigenvalues of~$\widehat{O}_0$ and analytic everywhere.
The logarithm of the ratio of the determinants then takes the following form
\begin{align}
	\label{eqn:det_ratio_as_lim_diff}
	\log \frac{\det \widehat{O}}{\det \widehat{O}_0} = \: & - \zeta_{\widehat{O}}'(0) + \zeta_{\widehat{O}_0}'(0) \notag \\
	= \: & - \lim_{t\to0} 
	\Bigg\{ e^{2\pi i t} \int_0^{e^{i\theta}\infty} d\lambda \, \left[ h_{\widehat{O}}(t;\lambda) - h_{\widehat{O}_0}(t;\lambda) \right] \notag \\
	&  \qquad - \frac{e^{2\pi i t} - 1}{2\pi i} \int_0^{e^{i\theta}\infty} d\lambda \, \log\lambda \, \left[ h_{\widehat{O}}(t;\lambda) - h_{\widehat{O}_0}(t;\lambda) \right] \Bigg\} \,,
\end{align}
where we have used the fact that the phase that~$h_{\widehat{O}}$ picks up when going around the origin, as well as its derivative with respect to~$t$, are both independent of the operator~$\widehat{O}$ itself.

\noindent
To take the~$t\to0$ limit, we recognise
\begin{align}
	\lim_{t\to0} h_{\widehat{O}} (t;\lambda) = \: \frac{d}{d\lambda} \log F_{\widehat{O}} (\lambda) \,.
\end{align}
Next, we assume that the last integral in Eq.\,\eqref{eqn:det_ratio_as_lim_diff} converges for~$t\to0$. In particular, this is equivalent to requiring that
\begin{equation}
	\label{eqn:decay_of_F}
	\left| \frac{d}{d\lambda} \log \frac{F_{\widehat{O}} (\lambda)}{F_{\widehat{O}_0} (\lambda)} \right| \overset{|\lambda|\to\infty}{<} \: \frac{1}{|\lambda|\log|\lambda|} \,.
\end{equation}
This crucial assumption only holds because we are interested in a ratio of functional determinants, as we will see later with an explicit choice of~$F$.
With the assumption of Eq.\,\eqref{eqn:decay_of_F}, the last term in Eq.\,\eqref{eqn:det_ratio_as_lim_diff} drops out in the limit~$t\to0$, and we are left with
\begin{align}
	\label{eqn:det_ratio_as_int}
	\log \frac{\det \widehat{O}}{\det \widehat{O}_0} = \: & \int_0^{e^{i\theta}\infty} d\lambda \, \frac{d}{d\lambda} \log \frac{F_{\widehat{O}} (\lambda)}{F_{\widehat{O}_0} (\lambda)}
	= \: \log \frac{F_{\widehat{O}}(0)}{F_{\widehat{O}_0}(0)} - \log \frac{F_{\widehat{O}}(e^{i\theta}\infty)}{F_{\widehat{O}_0}(e^{i\theta}\infty)} \,.
\end{align}
At this point, we need one last assumption, namely that the behaviour of~$F_{\widehat{O}}$ at complex infinity on the contour~$C_-$ is the same as that of~$F_{\widehat{O}_0}$, or in equations
\begin{equation}
	\label{eqn:behaviour_at_infty_F}
	\left| F_{\widehat{O}}(t e^{i\theta}) - F_{\widehat{O}_0}(t e^{i\theta}) \right| \xrightarrow{t\to\infty}\: 0 \,.
\end{equation}
With this, we find a final formula for the ratio of determinants in terms of the yet to be defined function~$F_{\widehat{O}}$, namely
\begin{equation}
	\label{eqn:det_ratio_as_F}
	\frac{\det \widehat{O}}{\det \widehat{O}_0} = \: \frac{F_{\widehat{O}}(0)}{F_{\widehat{O}_0}(0)} \,.
\end{equation}
Let us recall here the properties that~$F_{\widehat{O}}$ and~$F_{\widehat{O}_0}$ must satisfy for the derivation to go through:
\begin{itemize}
	\item They must be analytic over the whole complex plane,
	\item They must have simple zeroes at the eigenvalues of the associated operators and otherwise be non-vanishing,
	\item They must approach each other at infinity on the contour~$C_-$, namely they must satisfy Eq.\,\eqref{eqn:behaviour_at_infty_F},
	\item Their derivatives also must converge to one another and do so fast enough, namely they must satisfy Eq.\,\eqref{eqn:decay_of_F}.
\end{itemize}
Note that~$\theta$ is still a free parameter in this derivation, and we can choose it conveniently so that the last two requirements are satisfied.

Thanks to Eq.\,\eqref{eqn:det_ratio_as_F}, we only need to define an appropriate function~$F$ for the operators~$\widehat{O}$ and~$\widehat{O}_0$, and the ratio of their determinants is quickly found.
The question then becomes: what is a good choice of~$F$?
Different choices for this function are possible, leading to different formulas for the ratio of determinants.
One such choice leads to the Gel'fand-Yaglom theorem, and another yields the Green's function method.
We start with the former.

\subsection{The Gel'fand-Yaglom theorem}
\label{sec:gelfand}
\begin{figure}
    \centering
    \includegraphics[width=0.5\textwidth]{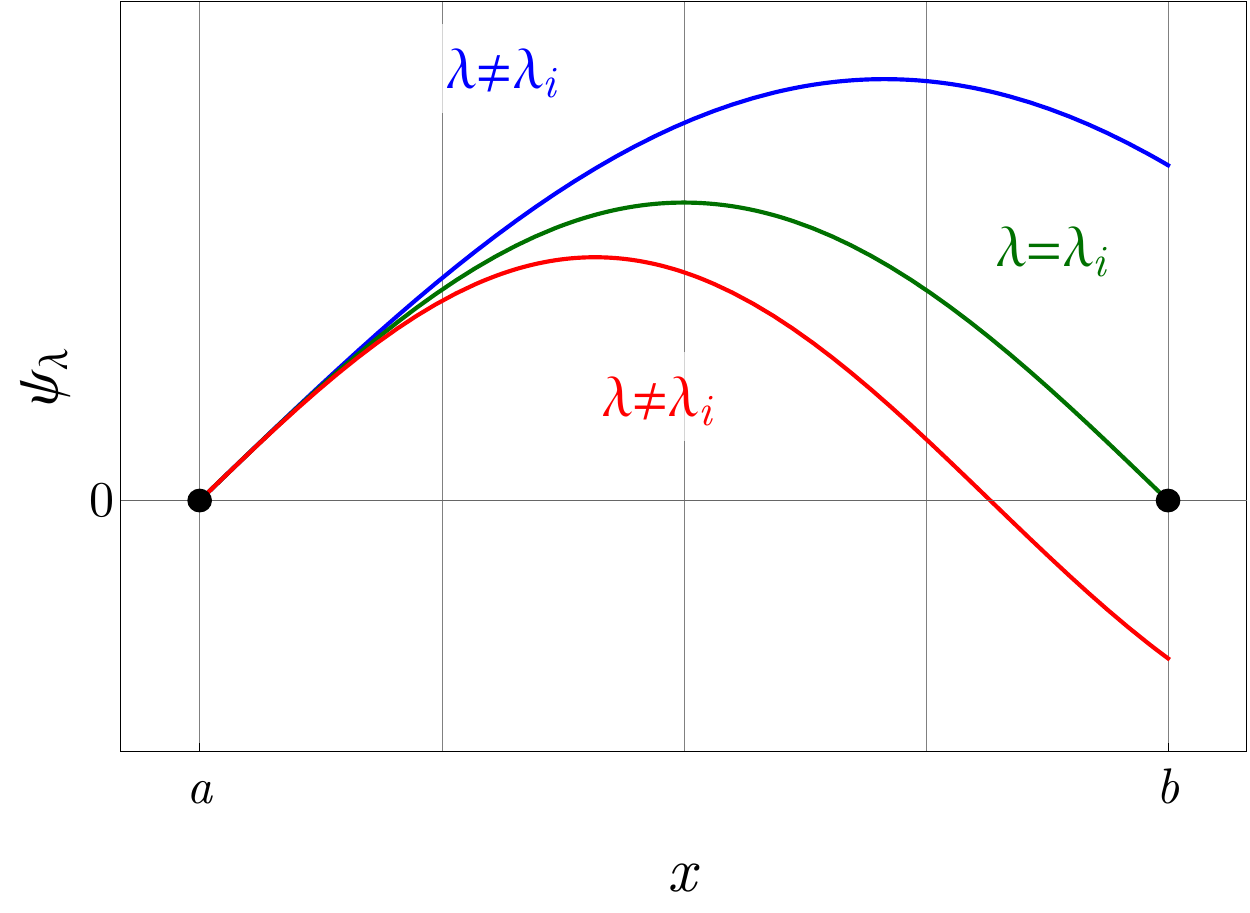}
    \caption{The function $\psi_{\widehat{O},\lambda}(x)$ solution to the differential problem of Eq.\,\eqref{eqn:GY_differential_problem} for some values of $\lambda$. When $\lambda$ is an eigenvalue of $\widehat{O}$, the function vanishes at the right boundary and coincides with the eigenfunction $\phi_{\widehat{O},i}$ solution to the eigenproblem of Eq.\,\eqref{eqn:eigenvalue_problem}.}
    \label{fig:GY_eigenproblem}
\end{figure}

Consider the following differential problem
\begin{equation}
	\label{eqn:GY_differential_problem}
	\begin{cases}
		(\widehat{O} \psi_{\widehat{O},\lambda} ) (x) = \: \lambda \psi_{\widehat{O},\lambda} (x) \\
		\psi_{\widehat{O},\lambda} (a) = \: 0 \,, \quad \psi'_{\widehat{O},\lambda}(a) = \: 1
	\end{cases}
	\,.
\end{equation}
This is a modified version of the eigenvalue problem of the operator~$\widehat{O}$ in Eq.\,\eqref{eqn:eigenvalue_problem}, with the crucial difference that it has Cauchy boundary conditions instead.
The differential problem in Eq.\,\eqref{eqn:GY_differential_problem} is well defined on~$\mathcal{H}$, so that it admits a solution $\psi_{\widehat{O},\lambda}$ for any complex $\lambda$, and the function $\psi_{\widehat{O},\lambda}$ is continuous on the interval $[a,b]$.
Let us define 
\begin{equation}
	F^{\mathrm{GY}}_{\widehat{O}} (\lambda) = \: \psi_{\widehat{O},\lambda} (b)\,,
\end{equation}
and analogously~$F^{\mathrm{GY}}_{\widehat{O}_0}$.
Given that $\psi_{\widehat{O},\lambda}$ is continuous in $[a,b]$, the function~$F^{\mathrm{GY}}_{\widehat{O}}$ is analytic over the whole complex plane.
Furthermore, when $\lambda=\lambda_i$ is one of the eigenvalues of~$\widehat{O}$, the function $\psi_{\widehat{O},\lambda}$ also solves the eigenvalue problem in Eq.\,\eqref{eqn:eigenvalue_problem} and it corresponds to the eigenfunction $\phi_{\widehat{O},i}$.
In particular, it vanishes at $x=b$, so that
\begin{align}
	F^{\mathrm{GY}}_{\widehat{O}} (\lambda_{\widehat{O},i}) = \: \phi_{\widehat{O},i}(b) = \: 0 \qquad \forall \lambda_{\widehat{O},i} \in \sigma(\widehat{O},\mathcal{H}) \,.
\end{align}
Analogously,~$F^{\mathrm{GY}}_{\widehat{O}_0}$ vanishes on the eigenvalues of~$\widehat{O}_0$.
This is exemplified in Figure~\ref{fig:GY_eigenproblem}, where the function $\psi_{\widehat{O},\lambda}$ is plotted for some generic differential operator $\widehat{O}$. For a generic value of $\lambda$, $\psi_{\widehat{O},\lambda}(b)\neq0$ acquires some finite value. However, when $\lambda=\lambda_i$ is in the spectrum of the differential operator, the function vanishes at the right boundary.

With this definition,~$F^{\mathrm{GY}}_{\widehat{O}}$ and~$F^{\mathrm{GY}}_{\widehat{O}_0}$ satisfy two of the four required properties.
To show that the other two properties are also satisfied, we have to specify something more about the operators~$\widehat{O}$ and~$\widehat{O}_0$.
We assume that the two operators $\widehat{O}$ and $\widehat{O}_0$ agree up to a function, namely
\begin{equation}
	\widehat{O} = \: \widehat{O}_0 + V\,,
\end{equation}
where~$V$ is the multiplication operator
\begin{equation}
	(V f)(x) = \: V(x) f(x)  \qquad \forall f\in \mathcal{H} \,.
\end{equation}
The function~$V(x)$ is analytic and bounded over the whole domain~$[a,b]$.
The boundedness of~$V(x)$ ensures that~$F^{\mathrm{GY}}_{\widehat{O}}$ and~$F^{\mathrm{GY}}_{\widehat{O}_0}$ satisfy the third and fourth properties, where the fourth property can be shown, for example, using the WKB approximation.

With this choice, we obtain the famous Gel'fand-Yaglom theorem, which in equations reads
\begin{equation}
	\label{eqn:GY_formula}
	\frac{\det\widehat{O}}{\det\widehat{O}_0} = \: \frac{\psi_{\widehat{O},0}(b)}{\psi_{\widehat{O}_0,0}(b)} \,.
\end{equation}
This formula is remarkably simple, and for this reason, it is widely used in the literature, both analytically and numerically, when computing ratios of determinants.
The numerical package \texttt{BubbleDet}\,\cite{Ekstedt_2023} uses the Gel'fand-Yaglom theorem for computing the ratio of functional determinants relevant for vacuum decay, and the same theorem has been applied for assessing the stability of the Standard Model\,\cite{Isidori_2001,Degrassi_2012,Andreassen_2018,Chigusa:2018uuj}.

Although in Eq.\,\eqref{eqn:eigenvalue_problem} and in the rest of this section we have been working with Dirichlet boundary conditions, the derivation of the theorem can easily be extended to more general Robin boundary conditions\,\cite{Kirsten:2004qv}. 
In particular, the boundary conditions relevant for vacuum decay are~$\phi_i'(a)=\phi_i(b)=0$.
Then, the Gel'fand-Yaglom theorem can be proven by modifying the boundary conditions in Eq.\,\eqref{eqn:GY_differential_problem}, so that the derivative of the wave function vanishes~$\psi'_{\widehat{O},\lambda}(a) = \: 0$ and its value is an arbitrary normalisation constant~$\psi_{\widehat{O},\lambda}(a) = \: 1$.
The final formula~\eqref{eqn:GY_formula} remains untouched.

The presence of zero modes affects the present discussion and, thus, the final formula, which needs to be modified accordingly.
This can be achieved by introducing a regulator that lifts the zero mode, allowing it to be extracted, as done in Ref.\,\cite{McKane:1995vp}. 
See Refs.\,\cite{Andreassen_2018,Chigusa:2018uuj,Ivanov:2022osf} for a modern presentation of the procedure.
A more sophisticated approach is developed in Ref.\,\cite{Kirsten:2003py}, where no regulator is introduced and instead one defines the RHS of Eq.\,\eqref{eqn:GY_formula} so that it is manifestly safe from the zero-mode problem.
In the present article, we refrain from presenting these well-established approaches and instead discuss the subtraction of the zero mode in the Green's function method in section~\ref{sec:zero-modes}.
First, we introduce the method itself.

\subsection{The Green's function method}\label{sec:resolvent_method}
The Green's function method, or resolvent method, is a tool equivalent to the Gel'fand-Yaglom theorem to compute ratios of functional determinants~\cite{Baacke:1993jr,Baacke:1993aj,Baacke:1994ix,Baacke:2008zx,Ai:2024taz}.
As we demonstrate in the following, it amounts to making a different choice for the function~$F$.
In section~\ref{sec:green_method}, we provide the original derivation of the method\,\cite{Baacke:1993aj}, which easily generalises to higher-dimensional operators.

Instead of specifying the function~$F$, we can directly choose the integrand of the contour integral in Eq.\,\eqref{eqn:zeta_as_contour_integral}.
An admissible choice is
\begin{equation}
	\label{eqn:F_as_resolvent}
	\frac{d}{d\lambda} \log F_{\widehat{O}}(\lambda) = \: - \int_a^b d\mu(x) \, G_{\widehat{O}}(-\lambda; x,x) \,,
\end{equation}
where~$G_{\widehat{O}}$ is the resolvent of the operator~$\widehat{O}$,
\begin{equation}
    \label{eqn:resolvent_defining_eq}
	\left[ \widehat{O}_x + s \right] G_{\widehat{O}}(s;x,y) = \: \delta_{(\mu)}(x-y) \,,
\end{equation}
and~$s$ is a constant of appropriate dimensionality.
Here, we have defined the $\delta$-function with respect to the measure $\mu$, namely
\begin{equation}
    \label{eqn:definition_delta_mu}
    \int_a^b d\mu(x)\, \delta_{(\mu)}(x-y) = \: 
    \begin{cases}
        1 \qquad \mathrm{if} \quad y \in [a,b] \,,\\
        0 \qquad \mathrm{otherwise} \,.
    \end{cases}
\end{equation}
As we see from Eq.\,\eqref{eqn:resolvent_defining_eq}, the resolvent $G_{\widehat{O}}$ is the Green's function of the operator $\widehat{O}+s$, hence the name of the method.
The integrand~$G_{\widehat{O}}(-\lambda;x,x)$ is analytic in the complex plane, except for having simple poles at the eigenvalues of the operator~$\widehat{O}$, as can be seen using the spectral representation,
\begin{equation}
	\label{eqn:spectral_resolvent}
	G_{\widehat{O}}(-\lambda;x,y) = \: \sum_{i} \frac{\phi_{\widehat{O},i}(x)\phi_{\widehat{O},i}^*(y)}{\lambda_{\widehat{O},i}-\lambda} \,,
\end{equation}
where the index $i$ spans over all the eigenvalues $\lambda_{\widehat{O},i}\in\sigma(\widehat{O},\mathcal{H})$.
The validity of the ansatz~\eqref{eqn:spectral_resolvent} can be readily checked by plugging it into Eq.\,\eqref{eqn:resolvent_defining_eq} for the resolvent, to find
\begin{align}
    \left[ \widehat{O}_x + s \right] G_{\widehat{O}}(s;x,y) = \: &
    \sum_i \underbrace{\left[ \widehat{O}_x + s \right] \phi_{\widehat{O},i}(x)}_{=(\lambda_{\widehat{O},i}+s)\phi_{\widehat{O},i}(x)} \frac{\phi_{\widehat{O},i}^*(y)}{\lambda_{\widehat{O},i}+s} \notag \\
    = \: & \sum_i \phi_{\widehat{O},i}(x) \phi_{\widehat{O},i}^*(y) = \: \delta_{(\mu)}(x-y) \,,
\end{align}
where in the last step we use the completeness relation of the eigenbasis.
From the spectral decomposition for $G_{\widehat{O}}(-\lambda;x,y)$, we see that the residue at each pole is~$-1$, and the derivation of the determinant goes through as before.
Back to the integral in Eq.\,\eqref{eqn:det_ratio_as_int}, we plug in our choice for the integrand, namely Eq.\,\eqref{eqn:F_as_resolvent},  and we find
\begin{align}
	\log \frac{\det \widehat{O}}{\det \widehat{O}_0}
	= \: & - \int_0^{e^{i\theta}\infty} d\lambda \, \int_a^b d\mu(x) \, \left[ G_{\widehat{O}}(-\lambda; x,x) - G_{\widehat{O}_0}(-\lambda;x,x) \right] \,.
\end{align}
Next, we can choose~$\theta=\pi$. In fact, even if the operator~$\widehat{O}$ has negative eigenvalues, we can define the integral around these poles with the principal value. Since all poles are simple, the principal value is well defined and always finite, as long as the integrand vanishes fast enough at infinity.
Choosing~$\theta=\pi$ and letting~$s=-\lambda$, we have the ratio of determinants in terms of the resolvent
\begin{equation}
	\label{eqn:resolvent_formula}
	\log \frac{\det \widehat{O}}{\det \widehat{O}_0} = \: - \int_0^\infty d s\, \int_a^b d \mu(x) \, \left[ G_{\widehat{O}}(s ; x,x) - G_{\widehat{O}_0}(s;x,x) \right] \,.
\end{equation}
The problem has been reduced to obtaining the Green's functions of the modified operators~$\widehat{O}+s$ and~$\widehat{O}_0+s$ respectively, and integrating the difference in the coincident limit.

The Green's function method is particularly convenient when we are interested in computing correlators around a non-trivial field configuration~$\varphi$.
Then, solving for the two-point correlator in the~$\varphi$ background means obtaining the Green's function~$G_{\varphi}$, which then straightforwardly leads to the resolvent for the operator $G_\varphi^{-1}$.
Additionally, the method can be easily generalised to compute the determinant of higher-dimensional operators, as we will see following its more conventional derivation in the next section.

\section{The Green's function method revisited}
\label{sec:green_method}
In the previous section, we provided a derivation of the Green's function method based on a contour integral argument. Historically, the method has instead been introduced via a spectral argument\,\cite{Baacke:1993aj,Baacke:1993jr}, which we now review.
Interestingly, the spectral argument can be made directly for a larger number of dimensions, where a generalisation of the contour integral procedure is less straightforward.

Let~$\mathcal{M}$ be a $D$-dimensional Euclidean manifold with coordinates~$\vec{x}$, and take~$\widehat{O}$ and~$\widehat{O}_0$ to be differential operators acting on a Hilbert space of functions from~$\mathcal{M}$ to the complex numbers.
The resolvent of operator~$\widehat{O}$ is defined by the differential equation
\begin{equation}	
	\label{eqn:resolvent_definition_multi_d}
	(\widehat{O}+s)\, G_{\widehat{O}}(s;\vec x, \vec y) = \: \delta^{(D)}_{(\mu)}(\vec x - \vec y) \,,
\end{equation}
and analogously for the resolvent~$G_{\widehat{O}_0}$.
Here, the $\delta$-function in higher dimensions is defined as the straightforward generalisation of Eq.\,\eqref{eqn:definition_delta_mu}.
Then, the formula for the multi-dimensional case reads
\begin{equation}
	\label{eqn:resolvent_formula_multi_d}
	\log \frac{\det \widehat{O}}{\det \widehat{O}_0} = \: - \int_0^\infty d s\, \int_{\mathcal{M}} d^D \!\mu(\vec{x}) \, \left[ G_{\widehat{O}}(s ; \vec x, \vec x) - G_{\widehat{O}_0}(s;\vec x, \vec x) \right] \,.
\end{equation}
To prove this, we observe that the resolvent as defined through Eq.\,\eqref{eqn:resolvent_definition_multi_d} can be written in the spectral decomposition as
\begin{equation}
	G_{\widehat{O}} (s;\vec x, \vec y) = \: \sum_i \frac{\phi_{\widehat{O},i}(\vec x) \phi_{\widehat{O},i}^*(\vec y)}{\lambda_{\widehat{O},i} + s} \,,
\end{equation}
where~$\lambda_{\widehat{O},i}\in\sigma(\widehat{O},\mathcal{H})$, and the eigenfunctions~$\{\phi_{\widehat{O},i}\}$ form an orthonormal eigenbasis of $\mathcal{H}$, namely
\begin{equation}
	\widehat{O} \phi_{\widehat{O},i}(\vec x) = \: \lambda_{\widehat{O},i} \phi_{\widehat{O},i}(\vec x) \,, \qquad \mathrm{and} \qquad \int_{\mathcal{M}} d\mu(\vec x) \, \phi_{\widehat{O},i}(\vec x) \phi_{\widehat{O},i'}^*(\vec x) = \: \delta_{ii'}\,.
\end{equation}
The resolvent of the operator~$\widehat{O}_0$ has an analogous spectral decomposition.
Then, plugging the spectral decomposition in the RHS of Eq.\,\eqref{eqn:resolvent_formula_multi_d}, we find
\begin{align}
	\log \frac{\det \widehat{O}}{\det \widehat{O}_0} = \: & - \int_0^\infty d s\, \int_{\mathcal{M}} d^D\mu(\vec x) \, \left[ \sum_i \frac{\phi_{\widehat{O},i}(\vec x) \phi_{\widehat{O},i}^*(\vec x)}{\lambda_{\widehat{O},i}+ s} - \sum_i \frac{\phi_{\widehat{O}_0,i}(\vec x) \phi_{\widehat{O}_0,i}^*(\vec x)}{\lambda_{\widehat{O}_0,i}+ s}  \right] \notag \\
	= \: & - \int_0^\infty d s\, \sum_i \left[ \frac{1}{\lambda_{\widehat{O},i}+ s} - \frac{1}{\lambda_{\widehat{O}_0,i}+ s}  \right] \notag \\
	= \: & \sum_i \log\frac{\lambda_{\widehat{O},i}}{\lambda_{\widehat{O}_0,i}} = \: \log \prod_i \frac{\lambda_{\widehat{O},i}}{\lambda_{\widehat{O}_0,i}} \,.
\end{align}
Note that in this derivation it is crucial to assume that the two operators~$\widehat{O}$ and~$\widehat{O}_0$ have the same number of eigenmodes for the $s$-integral to converge.
This assumption is not true when subtracting zero modes from one of the two operators, and more care is needed.
We describe the treatment of zero modes in section~\ref{sec:zero-modes}.

As a last comment, let us stress that, when $D>1$, the ratio of functional determinants is generally UV divergent~\cite{Ekstedt_2023,Baratella:2025dum}.
In the present method, the divergence arises from the coincident limit of the resolvent, $\lim_{\vec y\to\vec x}G(s;\vec x, \vec y)$, which is generally ill-defined.
A regularisation procedure is then required to render the ratio of determinants finite, and the contribution of counterterms is needed to renormalise the result.
A detailed discussion of the renormalisation of the ratio of functional determinants in the case of vacuum decay can be found in Ref.\,\cite{Carosi:2025jpe}.

\section{Zero modes}
\label{sec:zero-modes}
Our derivation for both the Gel'fand-Yaglom theorem and the Green's function method relied on the absence of zero modes.
At the end of section~\ref{sec:gelfand}, we highlighted that the Gel'fand-Yaglom theorem needs to be adapted when some eigenvalue vanishes and referred to previous work that tackles this aspect.
However, the literature lacks details on how to accommodate the presence of zero modes within the Green's function method.
In this section, we explore how the final formula~\eqref{eqn:resolvent_formula} must be modified to account for the presence of vanishing eigenvalues.
Once again, we work in one dimension and take coordinates on the interval $x\in[a,b]$.
We now assume that the operator~$\widehat{O}$ acting on the Hilbert space~$\mathcal{H}$ exhibits a zero mode, namely
\begin{equation}
	\widehat{O} \phi_0 = \: 0 \,, \qquad
	\phi_0 \in \mathcal{H}\,.
\end{equation}
Then, the resolvent of $\widehat{O}$ exhibits a pole for $s\to0$, which in turn makes the $s$-integral in Eq.\,\eqref{eqn:resolvent_formula_multi_d} divergent.
This is no error: the determinant vanishes because of the zero mode, and its logarithm blows up.
Generally, in physics, we are still interested in extracting the ratio of determinants after the zero eigenvalue is removed, for example, via the method of collective coordinates. 
This can be achieved rather straightforwardly if we can solve for (part of) the spectrum explicitly, as is the case when studying vacuum decay in the thin-wall limit, where the operator $\widehat{O}$ becomes of P\"oschl-Teller type, and its spectrum is known fully\,\cite{Dunne_2008,Marino:2015yie}.
Then, the zero mode can be subtracted from the calculation of the determinant by hand\,\cite{Lee:2014yud,Ivanov:2022osf}.
In the general case, however, we cannot solve for the spectrum, and we must define a procedure to subtract the zero mode when using the Green's function method.

The general strategy is to work in the space~$\mathcal{H}_\perp$ where the zero modes are absent, namely the subspace of~$\mathcal{H}$ which is orthogonal to~$\phi_0$.
We can easily build it by defining the projection operator~$\mathbb{P}_0$ by its action on a generic element~$\chi\in\mathcal{H}$,
\begin{equation}
	(\mathbb{P}_0 \chi)(x) = \: \phi_0(x) \int d\mu(y) \phi_0^*(y) \chi(y) \,.
\end{equation}
Then, the orthogonal component to~$\phi_0$ is given by
\begin{equation}
	\mathcal{H}_\perp = \: \left( \mathds{1} - \mathbb{P}_0 \right) \mathcal{H}\,.
\end{equation}
By construction, the operator~$\widehat{O}$ has no zero modes when acting on~$\mathcal{H}_\perp$. In particular, it can be inverted, and its resolvent satisfies a modified equation,
\begin{equation}
	\left( \widehat{O} + s \right)\, G_{\widehat{O}}^\perp(s) = \: \mathds{1}_\perp = \: \mathds{1} - \mathbb{P}_0 \,,
\end{equation}
where on the RHS we have the identity in the subspace~$\mathcal{H}_\perp$.
We use the apex~$\perp$ to highlight that~$G^\perp$ is the resolvent of~$\widehat{O}$ only on the subspace.
In coordinate space~$x,y\in[a,b]$, we have
\begin{equation}
	\label{eqn:resolvent_eqn_subtracted}
	\left[ \widehat{O} (x) + s \right] G_{\widehat{O}}^\perp (s; x,y) = \: \delta_{(\mu)}(x-y) - \phi_0^*(x) \phi_0(y)\,.
\end{equation}
At this point, we could plug the resolvent~$G_{\widehat{O}}^\perp$ in the formula~\eqref{eqn:resolvent_formula} to obtain the logarithm of the ratio of functional determinants once the zero eigenvalue is taken out.
Before doing that, let us stress that the resolvent equation is much easier to solve when the RHS only contains the~$\delta$-function, since it reduces to a Green's function equation for the operator~$\widehat{O}+s$.
It is convenient to define the resolvent in the full space~$\mathcal{H}$ as the solution to the equation
\begin{equation}
	\label{eqn:resolvent_eqn_full}
	\left[ \widehat{O} (x) + s \right] G_{\widehat{O}} (s; x,y) = \: \delta_{(\mu)}(x-y) \,.
\end{equation}
The resolvent on the subspace~$\mathcal{H}_\perp$ orthogonal to the zero mode can be obtained from it as
\begin{equation}
	\label{eqn:resolvent_subtracted}
	G_{\widehat{O}}^\perp (s; x,y) = \: G_{\widehat{O}} (s; x,y) - \frac{\phi_0^*(x) \phi_0(y)}{s} \,.
\end{equation}
Using Eq.\,\eqref{eqn:resolvent_eqn_full}, we can check that the ansatz in Eq.\,\eqref{eqn:resolvent_subtracted} solves Eq.\,\eqref{eqn:resolvent_eqn_subtracted}.
Thus, we can limit ourselves to finding the Green's function of the operator~$\widehat{O}+s$ on the full Hilbert space~$\mathcal{H}$ by solving Eq.\,\eqref{eqn:resolvent_eqn_full} with appropriate boundary conditions.
From the solution, we can obtain the resolvent on~$\mathcal{H}_\perp$ using Eq.\,\eqref{eqn:resolvent_subtracted}.
Strictly speaking, this procedure only works for~$s\neq0$.
However, this is sufficient, since we can define the resolvent at~$s=0$ as a limit of Eq.\,\eqref{eqn:resolvent_subtracted}, which is non-singular.

Now, back to computing the ratio of functional determinants of the operators~$\widehat{O}$ and~$\widehat{O}_0$.
We assume that the operator~$\widehat{O}_0$ has no zero modes, as is generally the case for most relevant scenarios.
Then, plugging the ansatz~\eqref{eqn:resolvent_subtracted} into the formula for the ratio of determinants~\eqref{eqn:resolvent_formula}, we have
\begin{align*}
	\log \frac{\sideset{}{'}\det \widehat{O}}{\det \widehat{O}_0} = \: & - \int_0^\infty d s \, \int_a^b d\mu(x) \, \left[ G_{\widehat{O}} ( s;x,x) - G_{\widehat{O}_0} (s;x,x) - \frac{\phi_0^*(x) \phi_0(x)}{s} \right] \,,
\end{align*}
where~$\sideset{}{'}\det$ means that the determinant is computed over the subspace~$\mathcal{H}_\perp$ where the zero mode~$\phi_0$ is absent.
This formula, however, has an issue.

Having subtracted a mode from $\det\widehat{O}$ but not from $\det\widehat{O}_0$, there is a mismatch in the number of modes.
As commented above, having the same number of modes is crucial for employing the Green's function method.
To fix this, we introduce a fictitious mode of eigenvalue~$m^2$ inside the integral, where~$m$ is an arbitrary number of appropriate dimensionality.
We must remember to remove this eigenvalue from the result after the integration is performed.
In practice, in the presence of a zero mode~$\phi_0$, we need to modify Eq.\,\eqref{eqn:resolvent_formula} as follows
\begin{align}
	\label{eqn:subtracted_resolvent_formula}
	\log \frac{\sideset{}{'}\det \widehat{O}}{\det \widehat{O}_0} = \: - \int_0^\infty d s\, \Bigg\{ & \int_a^b d \mu(x) \, \left[ G_{\widehat{O}}(s ; x,x) - G_{\widehat{O}_0}(s;x,x) - \frac{\phi_0^*(x) \phi_0(x)}{s} \right] \notag \\
	& + \frac{1}{m^2+s}\Bigg\} - \log m^2 \,.
\end{align}
We can extend the formula to the $D$-dimensional case in an analogous way as in Eq.\,\eqref{eqn:resolvent_formula_multi_d},
and it reads
\begin{align}
	\label{eqn:subtracted_resolvent_formula_multi_d}
	\log \frac{\sideset{}{'}\det \widehat{O}}{\det \widehat{O}_0} = \: - \int_0^\infty d s\, \Bigg\{ & \int_{\mathcal{M}} d^D\! \mu(\vec x) \, \left[ G_{\widehat{O}}(s ; \vec x,\vec x) - G_{\widehat{O}_0}(s;\vec x,\vec x) - \sum_{i_0=1}^{n_{0}}\frac{\phi_{i_0}^*(x) \phi_{i_0}(x)}{s} \right] \notag \\
	& + \frac{n_0}{m^2+s}\Bigg\} - n_0\log m^2 \,,
\end{align}
where~$n_0$ is the number of zero modes~$\phi_{i_0}$ of the operator~$\widehat{O}$.
There are several advantages to this subtraction procedure.
First of all, the Green's function method introduces a natural regulator for the zero mode, namely the parameter $s$. The vanishing eigenvalue leads to a pole in the resolvent as $s\to0$, which is naturally eliminated by subtracting the zero-mode contribution in the spectral decomposition.
Once this is done, we arrive at Eq.\,\eqref{eqn:subtracted_resolvent_formula_multi_d}, which is written fully in terms of convergent integrals, up to the presence of negative modes which we have not discussed yet.
Overall, we observe how the Green's function method accommodates a very natural treatment of the zero modes, whereas the Gel'fand-Yaglom theorem requires \emph{ad hoc} modifications\,\cite{Kirsten:2003py} or the introduction of a regulator\,\cite{McKane:1995vp}.

\subsection{About negative modes}
In carrying out the contour integral argument, we made only two critical assumptions on the spectrum of the operators: that it contained no zeroes and was bounded from below.
Negative modes are thus allowed, and a priori, both methods work just as well without modification.
When discussing the Green's function method, we highlighted how, in the presence of negative modes, the integral over the parameter~$s$ should be understood in the principal value sense.
This is because the resolvent~$G_{\widehat{O}}=(\widehat{O} + s)^{-1}$ has poles at~$s=-|\lambda_i|$ for each negative eigenvalue~$\lambda_i<0$.

Yet, when working numerically, principal value integrals are pretty nasty.
Instead, we can subtract the negative modes from the resolvent, just as we did for the zero modes, with the important difference that their eigenvalues should be included back in the final result.
To provide an updated formula for the ratio of determinants of operators~$\widehat{O}$ and~$\widehat{O}_0$, let us work directly in $D$ dimensions and assume that~$\widehat{O}$ has~$n_0$ zero modes~$\phi_{i_0}$ and~$n_{\mathrm{neg}}$ negative modes~$\phi_{i_\mathrm{neg}}$ with eigenvalues~$\lambda_{i_\mathrm{neg}}$, while~$\widehat{O}_0$ has neither.
Then, we can extend the formula~\eqref{eqn:subtracted_resolvent_formula_multi_d} in the presence of negative modes as follows
\begin{align}
	\label{eqn:subtracted_resolvent_formula_multi_d_no_neg}
	\log \frac{\sideset{}{'}\det \widehat{O}}{\det \widehat{O}_0} = \: & - \int_0^\infty d s\, \Bigg\{ \int_{\mathcal{M}} d^D\! \mu(\vec x) \, \Bigg[ G_{\widehat{O}}(s ; \vec x,\vec x) - G_{\widehat{O}_0}(s;\vec x,\vec x) - \sum_{i_0=1}^{n_{0}}\frac{\phi_{i_0}^*(\vec x) \phi_{i_0}(\vec x)}{s} \notag \\
	& - \sum_{i_{\mathrm{neg}}=1}^{n_{\mathrm{neg}}} \frac{\phi_{i_{\mathrm{neg}}}^*(\vec x) \phi_{i_{\mathrm{neg}}}(\vec x)}{\lambda_{i_{\mathrm{neg}}}+s} \Bigg]
	+ \frac{n_0+n_{\mathrm{neg}}}{m^2+s}\Bigg\} 
	- (n_0+n_{\mathrm{neg}}) \log m^2 + \sum_{i_{\mathrm{neg}}=1}^{n_{\mathrm{neg}}} \log \lambda_{i_{\mathrm{neg}}} \,.
\end{align}
Note that this requires knowledge about the negative eigenvalues and eigenfunctions, for which we must thus solve the eigenvalue problem, at least approximately.

\section{About the heat kernel method}
\label{sec:concerning_heat_kernel}
The heat kernel method is also a popular approach for computing functional determinants. It was first introduced in a field-theory context by Schwinger\,\cite{Schwinger:1951nm} and DeWitt\,\cite{DeWitt:1964mxt}, and it eventually became a standard tool for the calculation of effective actions\,\cite{Barvinsky:1985an,Vassilevich:2003xt}.
Its starting point is the definition of the heat kernel of the operator $\widehat{O}$ via the differential equation
\begin{equation}
    -\partial_\tau \mathcal{G}_{\widehat{O}}(\tau;x,y) = \: \widehat{O}_x \mathcal{G}_{\widehat{O}} (\tau;x,y)\,,
\end{equation}
with the initial condition $\mathcal{G}_{\widehat{O}}(0;x,y) = \delta(x-y)$.
This is formally solved by
\begin{equation}
    \label{eqn:heat_kernel_spectral_representation}
    \mathcal{G}_{\widehat{O}} (\tau;x,y) = \: \sum_i e^{-\lambda_i\tau} \phi_i(x) \phi_i^*(y) \,,
\end{equation}
where $\phi_i$ are the eigenfunctions with eigenvalue $\lambda_i$ of $\widehat{O}$.
Then, the $\zeta$-function of the operator $\widehat{O}$ can be written as
\begin{equation}
    \zeta_{\widehat{O}} (z) = \: \frac{1}{\Gamma(z)} \int_0^\infty d\tau \, \tau^{z-1} \int_a^b d x \, \mathcal{G}_{\widehat{O}} (\tau;x,x) \,,
\end{equation}
from which we can compute the determinant via Eq.\,\eqref{eqn: def det zeta function}.
The heat kernel method can be traced back to the Green's function method, and thus is fully equivalent to the Gel'fand-Yaglom theorem.
In fact, the resolvent function as defined in Eq.\,\eqref{eqn:resolvent_defining_eq} is nothing but the Laplace transform of the heat kernel, namely
\begin{equation}
    \label{eqn:resolvent_laplace_of_heat}
    G(s;x,y) = \: \int_0^\infty d\tau\, e^{-s\tau} \mathcal{G}(\tau;x,y) \,,
\end{equation}
which can be checked using Eq.\,\eqref{eqn:heat_kernel_spectral_representation} and~\eqref{eqn:spectral_resolvent}.
When plugging Eq.\,\eqref{eqn:resolvent_laplace_of_heat} into Eq.\,\eqref{eqn:resolvent_formula}, we find the ratio of determinants in terms of the heat kernel
\begin{align}
    \log \frac{\det \widehat{O}}{\det \widehat{O}_0} = \: & - \int_0^\infty d s\, \int_a^b d \mu(x) \, \int_0^\infty d\tau \, e^{-s\tau} \left[ \mathcal{G}_{\widehat{O}}(\tau ; x,x) - \mathcal{G}_{\widehat{O}_0}(\tau;x,x) \right] \notag \\
    = \: & - \int_0^\infty \frac{d\tau}{\tau} \int_a^b d \mu(x) \, \left[ \mathcal{G}_{\widehat{O}}(\tau ; x,x) - \mathcal{G}_{\widehat{O}_0}(\tau;x,x) \right] \,.
\end{align}
We can check the validity of this expression by using Eq.\,\eqref{eqn:heat_kernel_spectral_representation} to find
\begin{align}
    - \int_0^\infty \frac{d\tau}{\tau} \int_a^b d \mu(x) \, \left[ \mathcal{G}_{\widehat{O}}(\tau ; x,x) - \mathcal{G}_{\widehat{O}_0}(\tau;x,x) \right] = \: & - \sum_i \int_0^\infty \frac{d\tau}{\tau} \left[ e^{-\lambda_{\widehat{O},i}} - e^{-\lambda_{\widehat{O}_0,i}} \right] \notag \\
    = \: & \sum_i \log \frac{\lambda_{\widehat{O},i}}{\lambda_{\widehat{O}_0,i}} = \: \log \frac{\det \widehat{O}}{\det \widehat{O}_0} \,.
\end{align}
In summary, the heat-kernel method arises as an alternative formulation of the Green's function method through the Laplace transform.

\section{Conclusions}
\label{sec:conclusions}
In this article, we have provided a parallel derivation of the Gel'fand-Yaglom theorem and the Green's function method for computing the ratio of functional determinants, using a contour integral argument first introduced by Kirsten and McKane\,\cite{Kirsten:2003py}.
The key novelty of our work is having identified a different choice of integrand for the Kirsten-McKane argument, which naturally leads to the Green's function method as first introduced by Baacke and Junker\,\cite{Baacke:1993jr}.
For one-dimensional problems, the two approaches are completely equivalent in spirit and, generally, in complexity.

The Green's function method can be easily generalised to the higher-dimensional case, where its technical implementation can differ from the Gel'fand-Yaglom theorem, as observed by Baacke in Ref.\,\cite{Baacke:2008zx}.
After reviewing the conventional presentation of the method, we have clarified the treatment of zero and negative modes within the approach.
We have arrived at Eq.\,\eqref{eqn:subtracted_resolvent_formula_multi_d_no_neg}, a ready-to-use formula that represents a central result of this article.
Up to renormalisation, Eq.\,\eqref{eqn:subtracted_resolvent_formula_multi_d_no_neg} expresses the ratio of functional determinants fully in terms of finite quantities and convergent integrals, and it is thus particularly convenient for a numerical implementation.
For radially symmetric problems and Schrödinger-type operators, relevant to vacuum decay and bubble nucleation, the renormalisation of UV divergences is discussed in Ref.\,\cite{Carosi:2025jpe}.

Finally, we have recalled how the heat-kernel method is related to the Green's function method via a Laplace transform. Together with the rest of this article, this shows the complete equivalence of all three approaches, within the limits of validity of the contour-integral argument presented in section~\ref{sec:derivation}.

Having cleared the air about the validity of the various approaches, our work can serve as a basis for choosing the most convenient method, depending on one's needs.
The Gel'fand-Yaglom theorem is generally very powerful and has been streamlined for the most common applications, such as nucleation rates\,\cite{Ekstedt_2023}.
On the other hand, the Green's function method can be more appealing when the problem is inherently higher dimensional, such as for the nucleation of bubbles on domain walls\,\cite{Blasi:2022woz,Blasi:2023rqi} or strings\,\cite{Yajnik:1986tg,Blasi:2024mtc}. Also, it is of convenient use when we are also interested in computing perturbative quantities in a non-trivial background, which requires knowledge of the Green's function\,\cite{Carosi:2024lop}.

Next steps in our work include the release of publicly available code for the numerical implementation of the Green's function method for arbitrary one-dimensional operators, along with a comparison to the well-established \texttt{BubbleDet} package\,\cite{Ekstedt_2023}.

\acknowledgments
The author would like to thank Oliver Gould for his useful comments on an early version of this draft and for his encouragement to publish this material. The author is also grateful to Björn Garbrecht, Nils Wagner and Sara Maggio for fruitful discussions on these topics and helpful remarks on this draft.
The author's work is partially funded by the Fundamental Research Funds for the Central Universities in China (grant No. E5ER6601A2).

\bibliographystyle{JHEP}
\bibliography{literatur}

\end{document}